\documentclass[epjc]{webofc}

\usepackage[varg]{txfonts}   % \usepackage[utf8]{inputenc}
\usepackage{epsf}
\usepackage{latexsym,amssymb,euscript}
\usepackage{widetext}
\usepackage{amsmath}
\usepackage{nicefrac}
\usepackage{slashed} 
\usepackage{booktabs}
\usepackage{hyperref}
\usepackage{braket}
\usepackage{chngcntr}
\usepackage{bm}% bold math
\usepackage{bbold}
\usepackage{graphics}
\usepackage{graphicx}
\usepackage{mciteplus}
\usepackage{bbold}
\usepackage{caption}
\usepackage{subcaption}
\usepackage{pdfpages}
\usepackage[titletoc]{appendix}
\hypersetup{
 linktocpage = true,
 urlcolor = citegreen,
 colorlinks = true,
 linkcolor = citegreen,
 anchorcolor = citegreen,
 citecolor = purple,
 pdfstartview = {XYZ null null 1.25} 
           }
\usepackage{pstricks}
\usepackage{color}
\usepackage{xcolor}
\definecolor{urlblue}{rgb}{0.2,0.4,0.7}
\definecolor{citegreen}{rgb}{0,0.4,0.2}
\definecolor{linkred}{rgb}{0.9,0.2,0.1}
\usepackage{float}
\usepackage{academicons}
\definecolor{orcidlogocol}{HTML}{A6CE39}
\usepackage{fancyhdr}
\newcommand{\MSb}{\overline{\rm MS}}

\newcommand{\NLL}{{\rm NLL/NLO}}

\newcommand{\DY}{\Delta Y}

\newcommand{\Yps}{\Upsilon}

\newcommand{\tcite}[1]{~\cite{#1}}

\newcommand{\tarr}{
\begin{array}}
\newcommand{\earr}{\end{array}}

\begin{document}
\title{The high-energy limit of perturbative QCD: \\
Theory and phenomenology}
%
% subtitle is optional

\author{\firstname{Francesco Giovanni} \lastname{Celiberto}\inst{1,2,3}\fnsep\thanks{\email{fceliberto@ectstar.eu}} \and
        \firstname{Michael} \lastname{Fucilla}\inst{4,5}\fnsep\thanks{\email{michael.fucilla@unical.it}} \and
        \firstname{Alessandro} \lastname{Papa}\inst{4,5}\fnsep\thanks{\email{alessandro.papa@fis.unical.it} \, (corresponding author)}
        % etc.
}

\institute{ECT*, I-38123 Villazzano, Trento, Italy 
\and
Fondazione Bruno Kessler (FBK), I-38123 Povo, Trento, Italy
\and
INFN-TIFPA Trento Institute of Fundamental Physics and Applications, I-38123 Povo, Trento, Italy
\and
Universit\`a della Calabria, I-87036 Rende, Cosenza, Italy
\and
Istituto Nazionale di Fisica Nucleare, Gruppo collegato di Cosenza, I-87036 Rende, Cosenza, Italy
}

\abstract{%
  After a brief introduction of formal and phenomenological progresses in the study of the high-energy limit of perturbative QCD, we present arguments supporting the statement that the inclusive emission of Higgs bosons or heavy-flavored hadrons acts as fair stabilizer of high-energy resummed differential distributions.
  We come out with the message that the hybrid high-energy and collinear factorization, built in term of the next-to-leading logarithmic resummation \emph{\`a la} BFKL and supplemented by collinear parton distributions and fragmentation functions, is a valid and powerful tool to gauge the feasibility of precision analyses of QCD in its high-energy limit.
}
\maketitle
\section{Introductory remarks}
\label{introduction}

The Balitsky-Fadin-Kuraev-Lipatov (BFKL) resummation~\cite{Fadin:1975cb,Kuraev:1976ge,Kuraev:1977fs,Balitsky:1978ic} is the most powerful and adequate tool to account for large energy logarithms arising when QCD observables are studied in the Regge-Gribov or semi-hard regime\tcite{Gribov:1983ivg} (see\tcite{Celiberto:2017ius} for applications).
The BFKL approach permits us to resum leading logarithms (LL) $(\alpha_s\ln s)^n$, with $s$ is the center-of-mass energy squared and $\{Q\}$ one or a set of hard scales typical of the reaction, as well as next-to-leading ones (NLL), $\alpha_s^{n+1}\ln^n s$.
BFKL cross sections take the form of a high-energy convolution between a process-independent Green's function, which encodes the logarithmic resummation, and two impact factors, describing the fragmentation of each incoming particle. The BFKL Green's function evolves according to an integral equation, whose kernel is known within the next-to-leading order (NLO) for any fixed momentum transfer $t$ and for any possible two-gluon color exchange in the $t$-channel\tcite{Fadin:1998py,Ciafaloni:1998gs,Fadin:1998jv,Fadin:2000kx,Fadin:2000hu}.
Impact factors are process dependent and often represent the most challenging pieces in the calculation of the cross section.
They have been computed at the NLO for a very limited number of forward outgoing particles: (\emph{i}) quarks and gluons, (\emph{ii}) light jets, (\emph{iii}) light hadrons, (\emph{iv}) electroproduced vector mesons, (\emph{v}) photons, and (\emph{vi}) Higgs bosons in the infinite top-mass limit.
To test the BFKL resummation and, more in general, the high-energy dynamics of QCD at LHC energies and kinematic configurations, a \emph{hybrid} high-energy and collinear factorization was built\tcite{Colferai:2010wu,Bolognino:2018oth,Celiberto:2020tmb,Bolognino:2021mrc} (see\tcite{Deak:2009xt,Deak:2018obv,vanHameren:2022mtk} for a similar formalism) to encode collinear parton distribution functions (PDFs) and fragmentation functions (FFs) inside BFKL-factorized cross sections.
Probe channels for the hybrid factorization are: Mueller-Navelet jets\tcite{Ducloue:2013hia,Ducloue:2013bva,Caporale:2014gpa,Celiberto:2015yba,Celiberto:2015mpa,Celiberto:2016ygs,Celiberto:2016vva,Caporale:2018qnm,Celiberto:2022gji}, two-hadron\tcite{Celiberto:2016hae,Celiberto:2016zgb,Celiberto:2017ptm,Celiberto:2017uae,Celiberto:2017ydk}, hadron~$+$~jet\tcite{Bolognino:2018oth,Bolognino:2019cac,Bolognino:2019yqj,Celiberto:2020wpk,Celiberto:2020rxb,Celiberto:2022kxx}, Higgs~$+$~jet\tcite{Celiberto:2020tmb,Celiberto:2021fjf,Celiberto:2021tky,Celiberto:2021txb,Celiberto:2021xpm,Celiberto:2022fgx}, heavy-hadron~$+$~jet\tcite{Boussarie:2017oae,Celiberto:2017nyx,Bolognino:2019ouc,Bolognino:2019yls,Bolognino:2019ccd,Celiberto:2021dzy,Celiberto:2021fdp,Bolognino:2021zco,Bolognino:2022wgl,Celiberto:2022dyf,Celiberto:2022keu,Celiberto:2022zdg,Celiberto:2022kza}, heavy-light two-jet\tcite{Bolognino:2021mrc,Bolognino:2021hxxaux}, and multi-jet\tcite{Caporale:2015int,Caporale:2016soq,Caporale:2016xku,Celiberto:2016vhn,Caporale:2016zkc} inclusive tags.
In this work we present phenomenological arguments supporting the statement that large transverse masses observed in Higgs-boson emissions, as well as heavy-hadron detections, allow for clear stabilization of high-energy resummed observables under higher-order corrections and energy-scale variations.

\section{Natural stability}
\label{stability}

In this section we present phenomenological analyses to provide evidence that Higgs-boson emissions, as well as heavy-hadron detections, act as \textit{natural stabilizers} of the high-energy series. For our numerical analyses we use {\tt JETHAD}, a hybrid \textsc{ Python3.0/Fortran2008} modular package\tcite{Celiberto:2020wpk,Celiberto:2022rfj} aimed at the calculation, management, and processing of observables defined via distinct formalisms. Calculations are done in the $\MSb$ renormalization scheme and sensitivity to scale variation of our predictions has been evaluated by varying renormalization and factorization scales from half to two times their natural values suggested by kinematics. The error connected to the multi-dimensional integrations has constantly been kept below 1\%.

\subsection{Higgs + jet}
\label{ssec:Higgs}

We investigate the inclusive emission of a Higgs boson in association with a light-flavored jet at high rapidity distance, $\DY$. We investigate two different distributions: a) the total cross section differential in $\Delta Y$, and b) the Higgs transverse-momentum distribution at $\Delta Y=5$. 
The jet transverse momentum is in both cases integrated between 20 and 60 GeV, whereas the Higgs one is in the range 20 GeV and 2$M_t$ (with $M_t$ the top mass) in the first distribution, while it is instead left to vary in the same range in the second one. The jet rapidity lies in the range $y_J < |4.7|$ and the Higgs one in the $y_H < |2.5|$ range. The hadronic center-of-mass energy, $\sqrt s$, is fixed at 14 TeV.

\begin{figure}[t]
\centering

   \includegraphics[scale=0.33,clip]{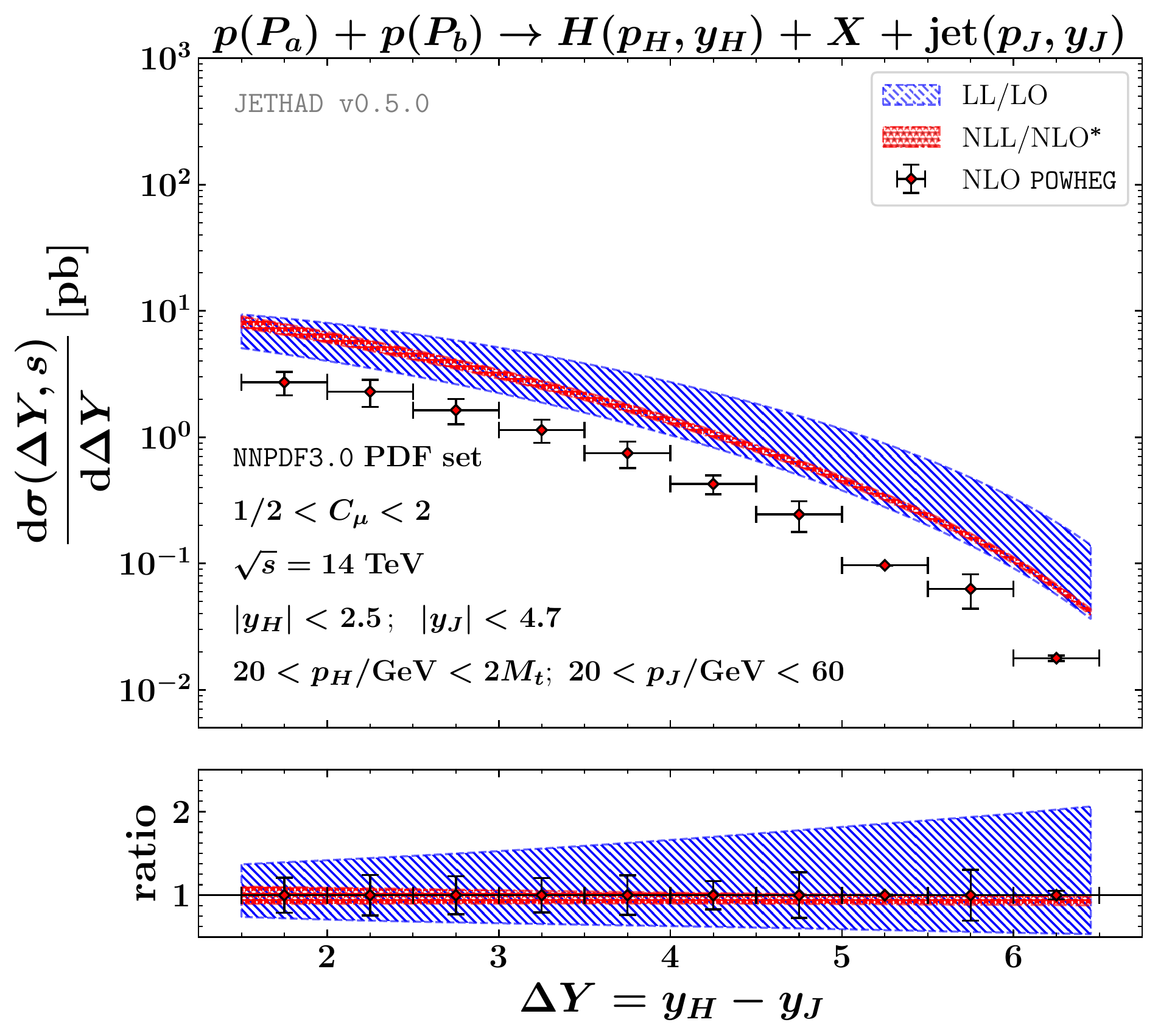}
%   \hspace{0.05cm}
   \includegraphics[scale=0.33,clip]{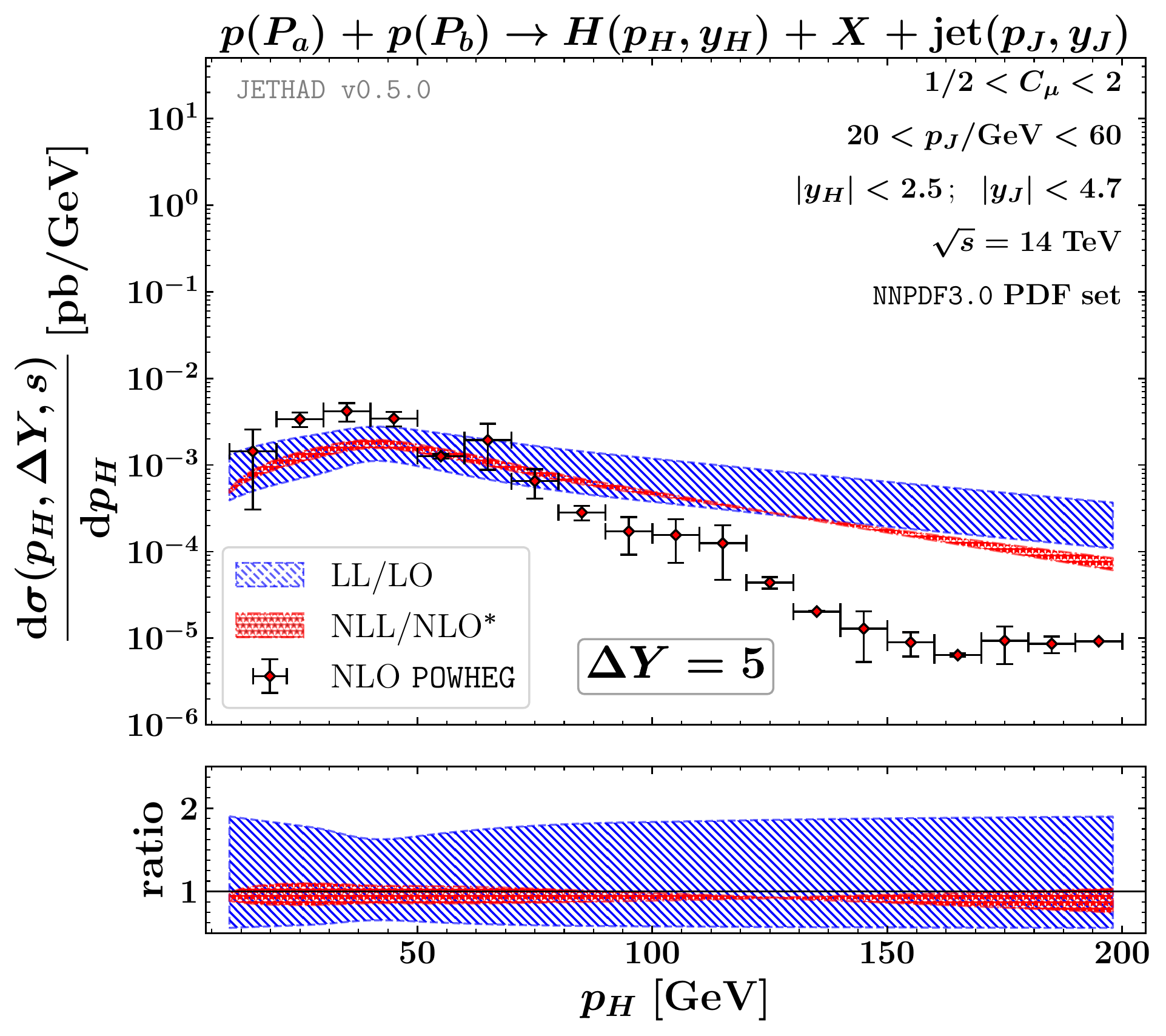}

\caption{Rapidity (left) and transverse-momentum distribution for the inclusive Higgs~$+$~jet production at 14~TeV LHC. Proton PDFs are described via the {\tt NNPDF4.0} NLO set\tcite{NNPDF:2021uiq,NNPDF:2021njg}}
\label{fig:Higgs}
\end{figure}

The total cross section is shown in the left panel of Fig.~\ref{fig:Higgs}. Different curves are for: the LL-resummed prediction (blue), the partially NLL-resummed series (red), and the fixed-order NLO result, obtained via the {\tt POWHEG} method\tcite{Nason:2004rx,Frixione:2007vw,Alioli:2010xd,Campbell:2012am} (bars).
The cross section shows a decreasing trend as the rapidity interval increases. This behavior is the net result of two opposite effects in the hybrid factorization.
On the one hand, the BFKL resummed partonic cross section grows as the squared partonic center-of-mass energy, $\hat{s}$, increases. Since $s$ is fixed, this corresponds to saying that the partonic cross section grows with increasing $\DY$, as expected. For the same reason, however, going to larger values of $\hat{s}$ means exploring kinematic regions associated with increasing value of the product $x_1 x_2$, where $x_1$ and $x_2$ are the Bjorken variables which characterize the two PDFs describing the incoming protons. The rapid falloff of PDFs in these kinematic regimes dominates and generates the downward trend. 
Notably, the NLL prediction has a narrower uncertainty band, which is always contained within the LL one. This represents a clear signal of the achieved stability at natural scales. Remarkably, fixed-order results are systematically lower than the resummed ones.

In the right panel of Fig.~\ref{fig:Higgs} the Higgs transverse-momentum distribution is shown. We distinguish three kinematic subregions:
\begin{itemize}
    \item The low-$p_T$ region, where $|\vec{p}_H|<10$ GeV; since it is dominated by large transverse-momentum logarithms, neglected by our formalism, we excluded it from our analysis;
    \item The intermediate-$p_T$ region, namely when $|\vec{p}_H| \sim |\vec{p}_J|$; this is the region where our formalism is expected to be adequate and we observe an impressive stability of the perturbative series;
    \item The large-$p_T$ region, which is essentially the last part of the tail; here DGLAP-type and threshold-type logarithms are relevant and our formalism is not the most appropriate one. 
\end{itemize}
We conclude from this analysis that the presence of a large energy-scale (the mass of the Higgs boson) improves the stability of the BFKL series.

\subsection{Heavy-flavored emissions}
\label{ssec:heavy-flavor}

We analyze total cross sections of two processes involving heavy-flavored emissions. It is important to specify that the presented studies refer to the production of heavy-flavored hadrons at large transverse momentum, a kinematic condition which allows us to adopt the variable-flavor number-scheme (VFNS)\tcite{Mele:1990cw,Cacciari:1993mq}.

\begin{figure}[t]
\centering

   \includegraphics[scale=0.33,clip]{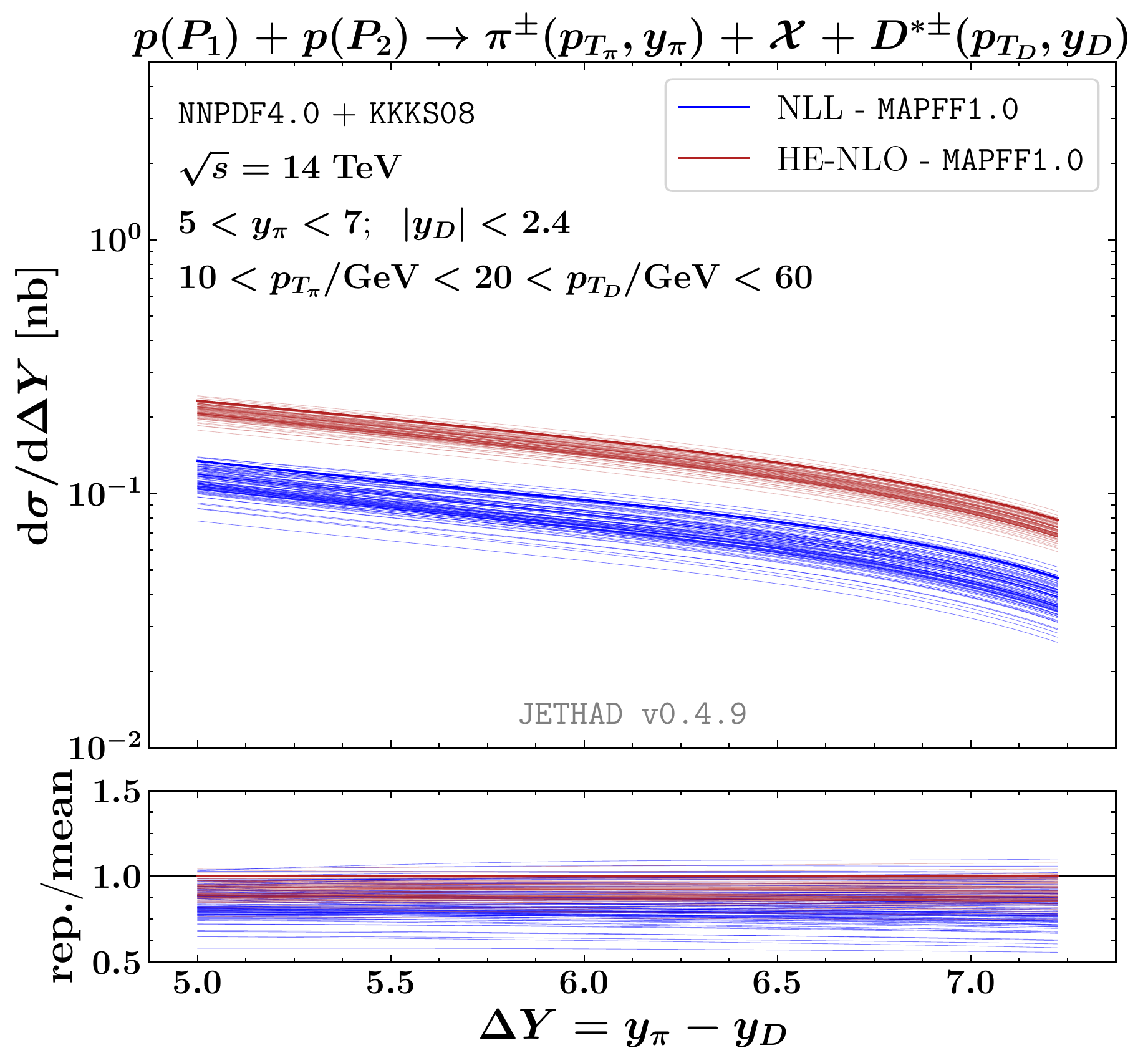}
%   \hspace{0.05cm}
   \includegraphics[scale=0.33,clip]{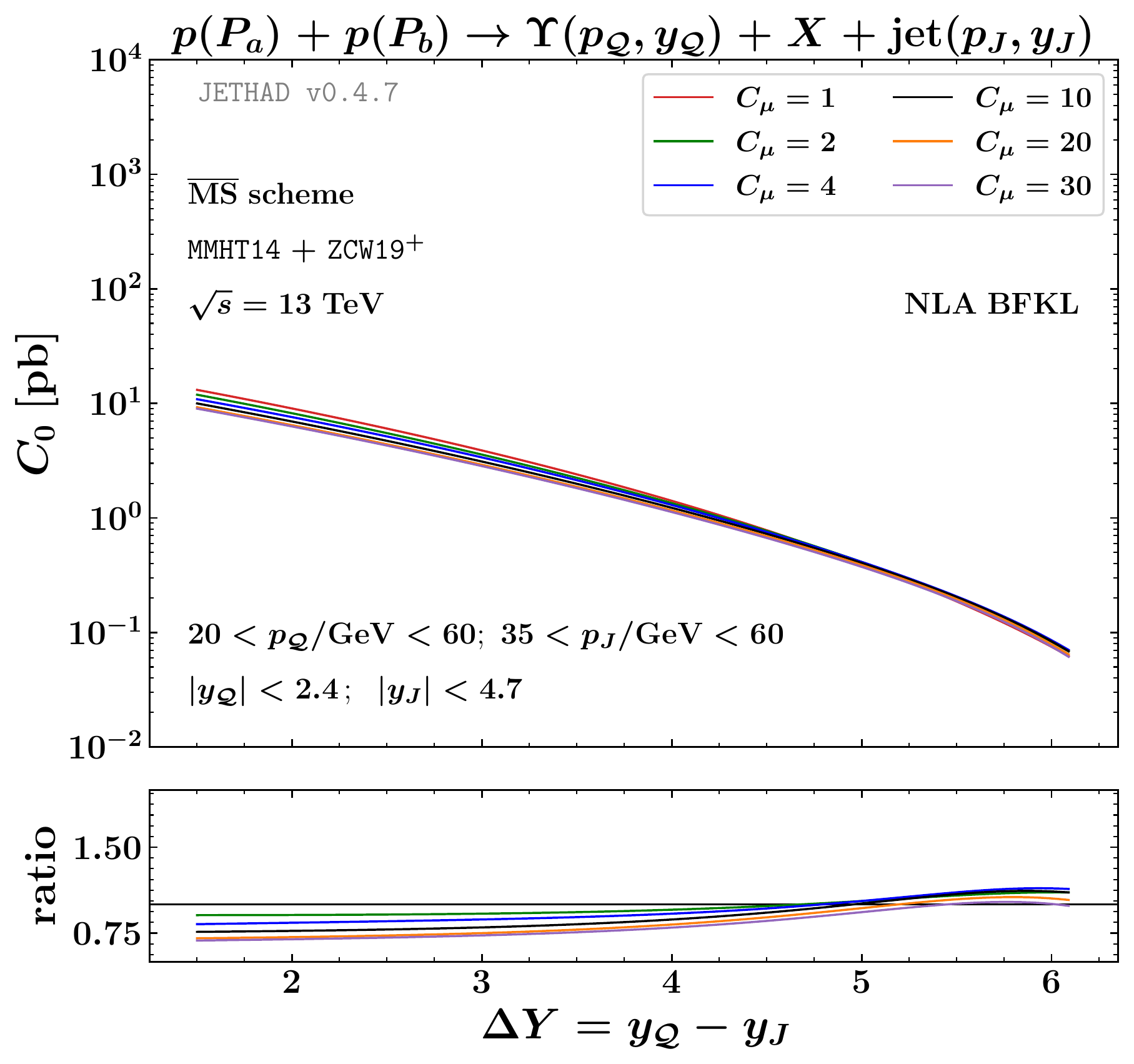}

\caption{Left panel: rapidity distribution for the inclusive $\pi^\pm$~$+$~$D^{*\pm}$ at 14~TeV FPF~$+$~ATLAS. {\tt NNPDF4.0} proton NLO PDFs\tcite{NNPDF:2021uiq,NNPDF:2021njg} are employed together with {\tt MAPFF1.0} pion NLO FFs\tcite{Khalek:2021gxf} and {\tt KKKS08} $D$-meson NLO FFs\tcite{Kneesch:2007ey}.
Right panel: rapidity distribution for the inclusive $\Yps$~$+$~jet channel at 13 TeV~LHC.
A study on progressive energy-scale variation in the range $1 < C_{\mu} < 30$ is made.
$\Yps$ fragmentation is described in terms of {\tt ZCW19$^+$} NLO FFs\tcite{Celiberto:2021dzy,Celiberto:2022kza,Zheng:2019dfk,Braaten:1993rw}}
\label{fig:heavy_flavor}
\end{figure}

The first reaction is the inclusive hadroproduction of a charged pion, $\pi^{\pm}$, detected at the planned Forward Physics Facility (FPF)\tcite{Anchordoqui:2021ghd,Feng:2022inv}, in association with a $D^{* \pm}$-meson tagged in the CMS barrel. Both particles feature large transverse momenta and are widely separated in rapidity. We choose as kinematic windows: $10 \; {\rm{GeV}} < |\vec p_{T_{\pi}}| < 20 \; {\rm{GeV}} < |\vec p_{T_{D}}| < 60 \; {\rm{GeV}} $, $5 < y_{\pi} < 7$ and $|y_D|<2.4$. 
In the left panel of Fig.~\ref{fig:heavy_flavor} we compare the NLL-resummed $\DY$-differential cross section (blue) with corresponding fixed-order NLO taken in the high-energy limit (blue). Different curves are obtained through the \textit{replica} method\tcite{Forte:2002fg,Ball:2021dab}. We observe that the weight of the resummation is quite large and the discrepancy between BFKL and fixed-order predictions is amplified by making use of asymmetric kinematic windows genuinely offered by a FPF~$+$~ATLAS coincidence setup\tcite{Celiberto:2022rfj,Celiberto:2022zdg}.

Finally, in the right panel of Fig.~\ref{fig:heavy_flavor} we present a study on the sensitivity on energy-scale variations of the $\DY$-distribution for the inclusive hadroproduction of a $\Yps$ in association with a light-flavored jet. The jet is tagged in kinematic configurations typical of current studies at the CMS detector\tcite{Khachatryan:2016udy}, namely~35~GeV~$< |\vec{p}_J| <$~60~GeV and $|y_J| < 4.7$. The quarkonium transverse momentum is chosen to be in the range 20~GeV~$< |\vec p_{\cal Q}| <$~60~GeV, in the spirit of a VFNS treatment. 
The $\Yps$ is detected by the CMS barrel detector, thus having $|y_{\cal Q}| < 2.4$. 
Predictions are stable even by varying scales by a factor 30 and stability improves as the $\DY$ increases.

We conclude this section by stressing the important difference between the two observed stability mechanisms. In the Higgs production case, the presence of a large energy scale, given by the Higgs transverse mass, provides us with the stabilization pattern.
Conversely, in the heavy-flavor production at large transverse momentum, NLO impact factors are calculated for light partons; here, the smooth and non-decreasing with $\mu_F$ behavior of the gluon FFs has a stabilizing effect on physical distributions.

\section{Future perspectives}
\label{conclusions}

We presented arguments supporting the statement that the inclusive emission of Higgs bosons or heavy-flavored bound states acts as fair stabilizer of high-energy resummed differential distributions.
We came out with the important message that the hybrid high-energy and collinear factorization, built in term of the NLL resummation \emph{\`a la} BFKL and supplemented by collinear PDFs and FF, is a valid and powerful tool to gauge the feasibility of precision analyses of QCD in its high-energy limit.

Future studies will extend this work to: $(i)$ full $\NLL$ analyses of Higgs-boson emissions by making use of the recently calculated NLO forward-Higgs impact factor\tcite{Celiberto:2022fgx,Hentschinski:2020tbi}, $(ii)$ the development of a \emph{multi-lateral} formalism in which distinct resummations are concurrently and consistently implemented, $(iii)$ accessing the low-$x$ proton structure at new-generation colliders\tcite{Chapon:2020heu,Anchordoqui:2021ghd,Feng:2022inv,Celiberto:2022rfj,Hentschinski:2022xnd,AbdulKhalek:2021gbh,Khalek:2022bzd,Acosta:2022ejc,AlexanderAryshev:2022pkx,Arbuzov:2020cqg,Amoroso:2022eow,Celiberto:2018hdy} via the connection of BFKL with unintegrated gluon densities\tcite{Hentschinski:2012kr,Bolognino:2018rhb,Bolognino:2018mlw,Bolognino:2019bko,Bolognino:2019pba,Celiberto:2019slj,Bolognino:2021niq,Bolognino:2021gjm,Bolognino:2022uty,Celiberto:2022fam,Bolognino:2022ndh,Motyka:2014lya,Celiberto:2018muu,Garcia:2019tne}, improved PDFs\tcite{Ball:2017otu,Bonvini:2019wxf} and polarized gluon transverse-momentum-dependent distributions\tcite{Bacchetta:2020vty,Celiberto:2021zww,Bacchetta:2021oht,Bacchetta:2021lvw,Bacchetta:2021twk,Bacchetta:2022esb,Bacchetta:2022crh,Bacchetta:2022nyv}.

\bibliography{references}

\end{document}